\documentclass[aps,prd,onecolumn,nofootinbib,superscriptaddress]{revtex4-2}

\usepackage{amsmath,amssymb,amsfonts}
\usepackage{bm}
\usepackage{graphicx}
\usepackage{xcolor}
\usepackage{hyperref}

\hypersetup{
    colorlinks=true,
    linkcolor=blue,
    citecolor=blue,
    urlcolor=blue
}

\newtheorem{theorem}{Theorem}
\newtheorem{proposition}{Proposition}

\begin{document}

\title{On the QCD Axion Potential in Fried's QCD Functional Formalism}

\author{Peter H. Tsang}
\email{phtsang@alum.mit.edu}

\date{\today}

\begin{abstract}
We examine the representation of the QCD axion potential in the context of Fried's nonperturbative QCD functional formalism. The axion field is introduced in the standard way by promoting the strong-CP angle to
\[
    \Theta=\theta_{\rm QCD}+\frac{a}{f_a}.
\]
The question addressed here is how the resulting \(\Theta\)-dependence of the QCD vacuum energy is encoded after the effective-locality reduction of the gluonic degrees of freedom.

The analysis is organized around two nonperturbative quantities. The first is the Fried chiral condensate,
\[
    \Sigma_{\rm F}=-\langle \bar q q\rangle_{\rm F},
\]
generated by the scalar/pseudoscalar projection of the effective-locality kernel. The second is the Fried pure-glue topological stiffness,
\[
    A_{\rm F}=\chi_{\rm YM}^{\rm F},
\]
which is represented in the Halpern formulation by a CP-odd self-dual/anti-self-dual variance.

Under these assumptions, the QCD topological susceptibility takes the interpolation form
\[
    \chi_{\rm top}^{\rm F}
    =
    \left[
    \frac{1}{A_{\rm F}}
    +
    \sum_f
    \frac{1}{m_f\Sigma_{\rm F}}
    \right]^{-1}.
\]
Consequently,
\[
    m_a^2 f_a^2=\chi_{\rm top}^{\rm F}.
\]
This expression has the expected limits: it reduces to \(A_{\rm F}\) in the heavy-quark limit, to
\[
    \Sigma_{\rm F}
    \left(
    \sum_f m_f^{-1}
    \right)^{-1}
\]
in the light-quark regime, and vanishes if any one quark mass is taken to zero.

In a separable scalar/pseudoscalar approximation to the effective-locality kernel, the condensate is written as
\[
    \Sigma_{\rm F}
    =
    \frac{N_c r\Lambda_{\rm EL}^3}{4\pi^2}I(r),
    \qquad
    r=\frac{M_0}{\Lambda_{\rm EL}},
\]
where \(r\) is determined by the gap equation
\[
    1=\alpha_\chi^{\rm F}J(r).
\]
The pure-glue stiffness is represented as
\[
    A_{\rm F}
    =
    \Lambda_{\rm EL}^4
    \left(
    \frac{g^2}{32\pi^2}
    \right)^2
    {\rm Var}_{\rm EL}(X_+-X_-),
\]
where \(X_+\) and \(X_-\) are the self-dual and anti-self-dual Halpern-field norms.

Thus the axion potential is reduced to two Fried-formalism objects: a chiral condensate and a CP-odd pure-glue variance. The result is conditional; a complete first-principles derivation requires the explicit evaluation of \(\Sigma_{\rm F}\) and \(A_{\rm F}\) from the full Fried--Gabellini--Grandou--Tsang--Sheu measure. We also show that the Fried-QCD contribution to a multi-axion mass matrix is rank one, so QCD alone generates only one massive axion direction. Additional axion-like species require additional independent topological sectors.
\end{abstract}

\maketitle
\tableofcontents

\section{Introduction}
\label{sec:introduction}

The QCD axion is usually introduced by promoting the strong-CP angle to a dynamical field,
\begin{equation}
    \theta_{\rm QCD}
    \longrightarrow
    \Theta(x)
    =
    \theta_{\rm QCD}+\frac{a(x)}{f_a}.
\end{equation}
The corresponding axion potential is determined by the \(\Theta\)-dependent part of the QCD vacuum energy,
\begin{equation}
    V_a(a)
    =
    E_{\rm vac}(\Theta)-E_{\rm vac}(0).
\end{equation}
For small \(\Theta\),
\begin{equation}
    E_{\rm vac}(\Theta)
    =
    E_{\rm vac}(0)
    +
    \frac12\chi_{\rm top}\Theta^2
    +
    O(\Theta^4),
\end{equation}
so that
\begin{equation}
    m_a^2f_a^2=\chi_{\rm top}.
\end{equation}
Here
\begin{equation}
    \chi_{\rm top}
    =
    \int d^4x\,
    \langle Q(x)Q(0)\rangle,
\end{equation}
with
\begin{equation}
    Q(x)
    =
    \frac{g^2}{32\pi^2}
    G^a_{\mu\nu}(x)\widetilde G^{a\mu\nu}(x).
\end{equation}

In standard treatments, \(\chi_{\rm top}\) may be computed or constrained using chiral perturbation theory, lattice QCD, instanton methods in the high-temperature regime, or phenomenological effective Lagrangians~\cite{GrilliDiCortona,GorghettoVilladoro,BorsanyiAxion}. The present paper does not modify the basic relation
\begin{equation}
    m_a^2f_a^2=\chi_{\rm top}.
\end{equation}
Instead, it asks how the same topological susceptibility is represented in Fried's nonperturbative QCD functional formalism~\cite{FriedReview,FriedPureGlue}.

Effective Locality is a nonperturbative property of QCD Green functions obtained after the gauge-invariant functional integration over gluonic degrees of freedom~\cite{FriedReview}. In this formulation, the resulting quark amplitudes are expressed through local color--Lorentz structures involving the Halpern field rather than through an expansion in ordinary perturbative gluon exchanges. This suggests a natural question: after the effective-locality reduction, what are the objects that determine the \(\Theta\)-dependence of the QCD vacuum energy?

The answer proposed here is conditional but explicit. The QCD axion potential is controlled by two Fried-formalism quantities. The first is the chiral condensate
\begin{equation}
    \Sigma_{\rm F}
    =
    -\langle\bar q q\rangle_{\rm F},
\end{equation}
generated by the scalar/pseudoscalar projection of the effective-locality kernel. The notation \(\Sigma_{\rm F}\) denotes the chiral condensate as obtained from the Fried Effective Locality kernel; in the present paper it is evaluated only in the scalar/pseudoscalar channel approximation.

The second is the pure Yang--Mills topological stiffness
\begin{equation}
    A_{\rm F}
    =
    \chi_{\rm YM}^{\rm F},
\end{equation}
represented in the Halpern sector by a CP-odd fluctuation of self-dual and anti-self-dual components. The notation \(A_{\rm F}\) is chosen to emphasize that this is the Fried-representation analogue of the pure Yang--Mills topological susceptibility, not an additional phenomenological parameter.

The central result is the interpolation formula
\begin{equation}
    \boxed{
    \chi_{\rm top}^{\rm F}
    =
    \left[
    \frac{1}{A_{\rm F}}
    +
    \sum_f
    \frac{1}{m_f\Sigma_{\rm F}}
    \right]^{-1}.
    }
    \label{eq:intro_interpolation}
\end{equation}
Consequently,
\begin{equation}
    \boxed{
    m_a^2f_a^2
    =
    \chi_{\rm top}^{\rm F}.
    }
\end{equation}

The terminology in this paper is deliberately conservative. We use ``Fried axion potential'' only as shorthand for the QCD axion potential after the QCD vacuum response has been represented in Fried's QCD functional framework, developed in the Fried--Gabellini--Grandou--Tsang--Sheu line of work, with Effective Locality as a central property. We do not mean that the axion field itself is generated by Effective Locality.

The paper is organized as follows. We first fix notation and introduce the \(\Theta\)-deformed QCD functional. We then derive the susceptibility interpolation formula. Next, we discuss the scalar/pseudoscalar effective-locality projection that generates \(\Sigma_{\rm F}\), and the Halpern representation of the pure-glue stiffness \(A_{\rm F}\). We then discuss the relation to the \(U(1)_A\) sector and the \(\eta'\), the phenomenological normalization of the QCD axion mass, and the rank-one structure of the QCD contribution to a multi-axion mass matrix. Finally, we state the limitations of the present result and the remaining Fried-formalism calculations required for a complete first-principles derivation.

\section{Notation and Conventions}
\label{sec:notation}

We use \(G^a_{\mu\nu}\) for the non-Abelian QCD field strength and
\begin{equation}
    \widetilde G^{a\mu\nu}
    =
    \frac12
    \epsilon^{\mu\nu\rho\sigma}
    G^a_{\rho\sigma}
\end{equation}
for its dual. The topological charge density is
\begin{equation}
    Q(x)
    =
    \frac{g^2}{32\pi^2}
    G^a_{\mu\nu}(x)\widetilde G^{a\mu\nu}(x).
\end{equation}
The axion-dependent QCD angle is denoted
\begin{equation}
    \Theta
    =
    \theta_{\rm QCD}+\frac{a}{f_a}.
\end{equation}
Throughout this paper, the subscript or superscript \({\rm F}\) denotes a quantity evaluated in Fried's QCD functional formalism, developed in the Fried--Gabellini--Grandou--Tsang--Sheu formulation of QCD. Thus
\begin{equation}
    \chi_{\rm top}^{\rm F}
\end{equation}
denotes the QCD topological susceptibility after the effective-locality reduction,
\begin{equation}
    \Sigma_{\rm F}
    =
    -\langle\bar q q\rangle_{\rm F}
\end{equation}
denotes the Fried chiral condensate, and
\begin{equation}
    A_{\rm F}
    =
    \chi_{\rm YM}^{\rm F}
\end{equation}
denotes the pure Yang--Mills topological stiffness in the Fried representation.

For three light flavors, \(f=u,d,s\), we define
\begin{equation}
    m_*
    =
    \left(
    \frac1{m_u}
    +
    \frac1{m_d}
    +
    \frac1{m_s}
    \right)^{-1}.
\end{equation}
The effective-locality scale in the scalar/pseudoscalar channel is denoted \(\Lambda_{\rm EL}\). The infrared dynamical mass generated by the Fried chiral gap equation is denoted \(M_0\), and
\begin{equation}
    r=\frac{M_0}{\Lambda_{\rm EL}}.
\end{equation}
The two dimensionless functions entering the chiral gap equation and condensate are
\begin{equation}
    J(r)
    =
    \int_0^\infty dx\,
    \frac{x e^{-2x}}
    {x+r^2e^{-2x}},
\end{equation}
and
\begin{equation}
    I(r)
    =
    \int_0^\infty dx\,
    \frac{x e^{-x}}
    {x+r^2e^{-2x}}.
\end{equation}
The Fried chiral gap equation is
\begin{equation}
    1=\alpha_\chi^{\rm F}J(r),
\end{equation}
where
\begin{equation}
    \alpha_\chi^{\rm F}
    =
    \frac{N_cN_fG_\chi\Lambda_{\rm EL}^2}{4\pi^2}.
\end{equation}
The corresponding chiral condensate is
\begin{equation}
    \Sigma_{\rm F}
    =
    \frac{N_cr\Lambda_{\rm EL}^3}{4\pi^2}I(r).
\end{equation}

The Halpern field is denoted \(\chi^a_{\mu\nu}\), and we introduce the dimensionless Halpern variable
\begin{equation}
    \chi^a_{\mu\nu}=\Lambda_{\rm EL}^2 y^a_{\mu\nu}.
\end{equation}
The self-dual and anti-self-dual parts of \(y^a_{\mu\nu}\) are
\begin{equation}
    y_{\mu\nu}^{a,\pm}
    =
    \frac12
    \left(
    y^a_{\mu\nu}
    \pm
    \widetilde y^a_{\mu\nu}
    \right).
\end{equation}
Their squared norms are denoted
\begin{equation}
    X_+=(y^+)^2,
    \qquad
    X_-=(y^-)^2.
\end{equation}
The dimensionless Fried pure-glue topological factor is
\begin{equation}
    {\cal A}_{\rm F}
    =
    \left(
    \frac{g^2}{32\pi^2}
    \right)^2
    {\rm Var}_{\rm EL}(X_+-X_-),
\end{equation}
so that
\begin{equation}
    A_{\rm F}=\Lambda_{\rm EL}^4{\cal A}_{\rm F}.
\end{equation}

With these definitions, the compact conditional Fried axion formula is
\begin{equation}
    \boxed{
    m_a^2f_a^2
    =
    \left[
    \frac{1}{\Lambda_{\rm EL}^4{\cal A}_{\rm F}}
    +
    \frac{4\pi^2}
    {N_cr\Lambda_{\rm EL}^3I(r)}
    \sum_f\frac1{m_f}
    \right]^{-1}.
    }
\end{equation}

\section{The \texorpdfstring{\(\Theta\)}{Theta}-Deformed QCD Functional}
\label{sec:theta_functional}

We begin with the QCD generating functional in the presence of a constant CP-odd source,
\begin{equation}
    Z(\Theta)
    =
    \int {\cal D}A\,{\cal D}q\,{\cal D}\bar q\,
    \exp\left\{
    iS_{\rm QCD}[A,q,\bar q]
    +
    i\Theta
    \int d^4x\,Q(x)
    \right\}.
    \label{eq:ZTheta}
\end{equation}
For the present discussion the axion field may be treated as slowly varying, so that the QCD vacuum energy may be regarded as a function of \(\Theta\). The topological susceptibility is
\begin{equation}
    \chi_{\rm top}
    =
    \left.
    \frac{\partial^2E_{\rm vac}(\Theta)}
    {\partial\Theta^2}
    \right|_{\Theta=0}.
\end{equation}

For light quarks, the \(\Theta\)-dependence may be moved from the topological term into the quark mass matrix by an anomalous \(U(1)_A\) rotation,
\begin{equation}
    M
    \longrightarrow
    M_\Theta
    =
    M\exp\left(
    i\gamma_5\frac{\Theta}{N_f}
    \right).
\end{equation}
At low energy the same content is encoded in the anomalous chiral potential
\begin{equation}
    V_{\rm F}
    =
    -\Sigma_{\rm F}
    \sum_f m_f\cos\phi_f
    +
    \frac{A_{\rm F}}{2}
    \left(
    \Theta-\sum_f\phi_f
    \right)^2.
    \label{eq:chiral_potential_functional_section}
\end{equation}
Here \(\phi_f\) are the light-quark chiral phases, \(\Sigma_{\rm F}\) is the Fried condensate, and \(A_{\rm F}\) is the pure Yang--Mills topological stiffness in the Fried representation.

The complementary representation keeps the gluonic topological source explicit. In the Halpern representation one introduces an antisymmetric auxiliary tensor field \(\chi^a_{\mu\nu}(x)\), and the local color--Lorentz matrix
\begin{equation}
    {\cal F}^{ab}_{\mu\nu}(\chi)=f^{abc}\chi^c_{\mu\nu}.
\end{equation}
The CP-odd density is represented by the corresponding local pseudoscalar Halpern invariant,
\begin{equation}
    G^a_{\mu\nu}\widetilde G^{a\mu\nu}
    \quad\leadsto\quad
    \chi^a_{\mu\nu}\widetilde\chi^{a\mu\nu}.
\end{equation}
This notation denotes the representation of the CP-odd insertion inside the Halpern/effective-locality functional. It is not a classical pointwise identification of \(G\) with \(\chi\).

In a local effective-locality cell approximation, the pure-glue \(\Theta\)-dependent factor may be written as
\begin{equation}
    z_{\rm F}(\Theta)
    =
    \int d\mu_{\rm EL}(y)\,
    \exp\left[
    i\Theta q_\chi(y)
    \right],
\end{equation}
where
\begin{equation}
    q_\chi(y)
    =
    \frac{g^2}{32\pi^2}
    y^a_{\mu\nu}\widetilde y^{a\mu\nu}.
\end{equation}
The corresponding pure-glue stiffness is
\begin{equation}
    A_{\rm F}
    =
    \Lambda_{\rm EL}^4
    \left[
    \langle q_\chi^2\rangle_{\rm EL}
    -
    \langle q_\chi\rangle_{\rm EL}^2
    \right].
\end{equation}
For a CP-even measure at \(\Theta=0\), \(\langle q_\chi\rangle_{\rm EL}=0\), so \(A_{\rm F}=\Lambda_{\rm EL}^4\langle q_\chi^2\rangle_{\rm EL}\).

The anomalous chiral representation and the Halpern representation isolate the two quantities that enter the axion susceptibility: \(\Sigma_{\rm F}\) and \(A_{\rm F}\).

\section{Main Results in Theorem Form}
\label{sec:main_results}

Before giving the derivations, we collect the main results in theorem form.

\begin{theorem}[Fried susceptibility interpolation]
\label{thm:fried_interpolation}
Assume that the Effective Locality reduction supplies a chiral condensate \(\Sigma_{\rm F}=-\langle\bar q q\rangle_{\rm F}\), and a pure Yang--Mills topological stiffness \(A_{\rm F}=\chi_{\rm YM}^{\rm F}\). Then the QCD topological susceptibility in the presence of light quarks is
\begin{equation}
    \boxed{
    \chi_{\rm top}^{\rm F}
    =
    \left[
    \frac{1}{A_{\rm F}}
    +
    \sum_f
    \frac{1}{m_f\Sigma_{\rm F}}
    \right]^{-1}.
    }
\end{equation}
Consequently, for the QCD axion,
\begin{equation}
    \boxed{m_a^2 f_a^2=\chi_{\rm top}^{\rm F}.}
\end{equation}
\end{theorem}

\begin{proposition}[Consistency limits]
\label{prop:consistency_limits}
The susceptibility has the correct pure-glue, light-quark, and massless-quark limits: it reduces to \(A_{\rm F}\) in the heavy-quark limit, to \(\Sigma_{\rm F}/\sum_f1/m_f\) in the light-quark regime, and to zero if any one quark mass is taken to zero.
\end{proposition}

\begin{proposition}[Effective Locality chiral condensate]
\label{prop:fried_condensate}
In the separable scalar/pseudoscalar approximation to the Effective Locality kernel, with \(M_{\rm F}(p^2)=M_0\exp(-p^2/\Lambda_{\rm EL}^2)\) and \(r=M_0/\Lambda_{\rm EL}\), the chiral gap equation is
\begin{equation}
    \boxed{1=\alpha_\chi^{\rm F}J(r),}
\end{equation}
with
\begin{equation}
    J(r)=\int_0^\infty dx\,\frac{x e^{-2x}}{x+r^2e^{-2x}}.
\end{equation}
The corresponding condensate is
\begin{equation}
    \boxed{
    \Sigma_{\rm F}=\frac{N_cr\Lambda_{\rm EL}^3}{4\pi^2}I(r),
    }
\end{equation}
where
\begin{equation}
    I(r)=\int_0^\infty dx\,\frac{x e^{-x}}{x+r^2e^{-2x}}.
\end{equation}
A nonzero chiral solution requires \(\alpha_\chi^{\rm F}>2\).
\end{proposition}

\begin{proposition}[Pure-glue Halpern variance]
\label{prop:halpern_variance}
Let \(y^a_{\mu\nu}\) be the dimensionless Halpern field, and let \(X_+\) and \(X_-\) be the norms of its self-dual and anti-self-dual components. Then the pure-glue Fried topological stiffness is represented as
\begin{equation}
    \boxed{
    A_{\rm F}
    =
    \Lambda_{\rm EL}^4
    \left(
    \frac{g^2}{32\pi^2}
    \right)^2
    {\rm Var}_{\rm EL}(X_+-X_-).
    }
\end{equation}
\end{proposition}

Combining Proposition~\ref{prop:fried_condensate} and Proposition~\ref{prop:halpern_variance} with Theorem~\ref{thm:fried_interpolation} gives
\begin{equation}
    \boxed{
    m_a^2f_a^2
    =
    \left[
    \frac{1}{\Lambda_{\rm EL}^4{\cal A}_{\rm F}}
    +
    \frac{4\pi^2}
    {N_cr\Lambda_{\rm EL}^3I(r)}
    \sum_f\frac1{m_f}
    \right]^{-1}.
    }
\end{equation}

\begin{theorem}[Rank-one QCD axion mass matrix]
\label{thm:rank_one_qcd}
Suppose \(N\) axion-like fields couple to QCD only through the single effective angle \(\Theta_Q=\theta_Q+\sum_i c_i a_i/f_i\). Then the Fried-QCD contribution to their mass matrix is
\begin{equation}
    \boxed{
    (M_Q^2)_{ij}
    =
    \chi_Q^{\rm F}
    \frac{c_ic_j}{f_if_j}.
    }
\end{equation}
This matrix has rank one. Therefore Fried-QCD alone generates only one massive QCD axion direction.
\end{theorem}

\section{The Susceptibility Interpolation Formula}
\label{sec:susceptibility_interpolation}

We now derive Theorem~\ref{thm:fried_interpolation}. The derivation uses only the anomalous chiral structure of QCD and the assumption that the Fried effective-locality reduction supplies \(\Sigma_{\rm F}\) and \(A_{\rm F}\).

The low-energy \(\Theta\)-dependent potential is written as
\begin{equation}
    V_{\rm F}
    =
    -\Sigma_{\rm F}
    \sum_f m_f\cos\phi_f
    +
    \frac{A_{\rm F}}{2}
    \left(
    \Theta-\sum_f\phi_f
    \right)^2 .
\end{equation}
For small \(\Theta\), expand \(\cos\phi_f=1-\phi_f^2/2+O(\phi_f^4)\). Dropping the constant part gives
\begin{equation}
    V_{\rm F}^{(2)}
    =
    \frac{\Sigma_{\rm F}}{2}
    \sum_f m_f\phi_f^2
    +
    \frac{A_{\rm F}}{2}
    \left(
    \Theta-\sum_f\phi_f
    \right)^2 .
\end{equation}
Define
\begin{equation}
    \lambda=\Theta-\sum_f\phi_f .
\end{equation}
Stationarity with respect to \(\phi_f\) gives
\begin{equation}
    \Sigma_{\rm F}m_f\phi_f-A_{\rm F}\lambda=0,
\end{equation}
so
\begin{equation}
    \phi_f=\frac{A_{\rm F}\lambda}{m_f\Sigma_{\rm F}}.
\end{equation}
Summing over flavors,
\begin{equation}
    \sum_f\phi_f
    =
    A_{\rm F}\lambda
    \sum_f\frac{1}{m_f\Sigma_{\rm F}}.
\end{equation}
Since \(\lambda=\Theta-\sum_f\phi_f\),
\begin{equation}
    \lambda
    \left[
    1+
    A_{\rm F}
    \sum_f\frac{1}{m_f\Sigma_{\rm F}}
    \right]
    =
    \Theta.
\end{equation}
Thus
\begin{equation}
    \lambda
    =
    \frac{\Theta}{
    1+
    A_{\rm F}
    \sum_f
    \dfrac{1}{m_f\Sigma_{\rm F}}
    }.
\end{equation}
At the stationary point,
\begin{equation}
    \frac{dV_{\rm eff}^{\rm F}}{d\Theta}=A_{\rm F}\lambda.
\end{equation}
Therefore
\begin{equation}
    \boxed{
    \chi_{\rm top}^{\rm F}
    =
    \frac{A_{\rm F}}{
    1+
    A_{\rm F}
    \sum_f
    \dfrac{1}{m_f\Sigma_{\rm F}}
    }
    =
    \left[
    \frac{1}{A_{\rm F}}
    +
    \sum_f
    \frac{1}{m_f\Sigma_{\rm F}}
    \right]^{-1}.
    }
\end{equation}

In the physical light-quark regime,
\begin{equation}
    \chi_{\rm top}^{\rm F}
    =
    m_*\Sigma_{\rm F}
    \left[
    1-
    \frac{m_*\Sigma_{\rm F}}{A_{\rm F}}
    +
    O\left(
    \frac{m_*^2\Sigma_{\rm F}^2}{A_{\rm F}^2}
    \right)
    \right].
\end{equation}
Thus the leading physical QCD axion mass is controlled by the Fried chiral condensate.

\section{Fried's Formalism Versus an Effective Model}
\label{sec:EL_vs_model}

Before deriving the scalar/pseudoscalar gap equation, it is important to clarify the status of the effective chiral interaction used below. The construction is not meant to introduce an independent low-energy four-fermion model. The starting point remains the QCD functional integral, including the CP-odd source \(\Theta\int d^4x\,Q(x)\).

In the Fried--Gabellini--Grandou--Tsang--Sheu formulation, the gluonic functional integration is carried out in such a way that quark Green functions are governed by local color--Lorentz structures involving the Halpern field. This is the phenomenon of Effective Locality. The scalar/pseudoscalar interaction
\begin{equation}
    {\cal L}_{\chi}^{\rm EL}
    =
    G_\chi
    \left[
    (\bar q q)^2
    +
    (\bar q i\gamma_5\vec\tau q)^2
    \right]
    +\cdots
\end{equation}
should therefore be read as a channel projection of the full effective-locality kernel. The coefficient \(G_\chi\) is not a fundamental parameter of a separate model. It denotes the scalar/pseudoscalar projection of the Fried kernel, \(K_{\rm EL}\to G_\chi\).

The present paper gives a reduction theorem: if the Fried formalism supplies \(\Sigma_{\rm F}\) and \(A_{\rm F}\), then the QCD axion potential follows from Eq.~\eqref{eq:intro_interpolation}.

\section{Scalar/Pseudoscalar Projection of the Effective-Locality Kernel}
\label{sec:chiral_projection}

After the gauge-invariant gluonic functional integration, quark currents interact through local color--Lorentz structures built from the Halpern field. In a schematic local approximation, the induced current-current interaction may be written as
\begin{equation}
    {\cal L}_{\rm EL}^{(4q)}
    =
    -G_{\rm EL}
    \left(
    \bar q\gamma_\mu T^a q
    \right)
    \left(
    \bar q\gamma^\mu T^a q
    \right),
\end{equation}
where \({\rm tr}(T^aT^b)=\delta^{ab}/2\). Projecting into the color-singlet scalar/pseudoscalar \(q\bar q\) channel gives
\begin{equation}
    {\cal L}_{\chi}^{\rm EL}
    =
    G_\chi
    \left[
    (\bar q q)^2
    +
    (\bar q i\gamma_5\vec\tau q)^2
    \right]
    +\cdots,
\end{equation}
where \(G_\chi=C_\chi G_{\rm EL}\) and
\begin{equation}
    C_\chi=\frac{N_c^2-1}{2N_c^2}.
\end{equation}
For QCD, \(C_\chi=4/9\).

To obtain an analytic expression, approximate the scalar/pseudoscalar kernel by
\begin{equation}
    G_\chi(p,k)=G_\chi f(p^2)f(k^2),
    \qquad
    f(p^2)=\exp(-p^2/\Lambda_{\rm EL}^2).
\end{equation}
The corresponding dynamical mass is
\begin{equation}
    M_{\rm F}(p^2)=M_0 f(p^2).
\end{equation}
The gap equation becomes
\begin{equation}
    1
    =
    4N_cN_fG_\chi
    \int\frac{d^4k}{(2\pi)^4}
    \frac{f^2(k^2)}{k^2+M_0^2f^2(k^2)}.
\end{equation}
Using \(r=M_0/\Lambda_{\rm EL}\), this is
\begin{equation}
    \boxed{1=\alpha_\chi^{\rm F}J(r),}
\end{equation}
where
\begin{equation}
    \alpha_\chi^{\rm F}=\frac{N_cN_fG_\chi\Lambda_{\rm EL}^2}{4\pi^2},
    \qquad
    J(r)=\int_0^\infty dx\,\frac{x e^{-2x}}{x+r^2e^{-2x}}.
\end{equation}
Since \(J(0)=1/2\), chiral symmetry breaking requires \(\alpha_\chi^{\rm F}>2\).

The condensate is
\begin{equation}
    \Sigma_{\rm F}
    =
    4N_c
    \int\frac{d^4p}{(2\pi)^4}
    \frac{M_{\rm F}(p^2)}{p^2+M_{\rm F}^2(p^2)},
\end{equation}
so
\begin{equation}
    \boxed{
    \Sigma_{\rm F}
    =
    \frac{N_cr\Lambda_{\rm EL}^3}{4\pi^2}I(r),
    }
\end{equation}
where
\begin{equation}
    I(r)=\int_0^\infty dx\,\frac{x e^{-x}}{x+r^2e^{-2x}}.
\end{equation}

Writing
\begin{equation}
    G_{\rm EL}
    =
    \frac{g^2\Delta_\chi^2}{\Lambda_{\rm EL}^2}
    \left[1+{\cal C}_{\rm loop}\right],
\end{equation}
one obtains, for \(N_c=N_f=3\),
\begin{equation}
    \boxed{
    \alpha_\chi^{\rm F}
    =
    \frac{g^2}{\pi^2}
    \Delta_\chi^2
    \left[1+{\cal C}_{\rm loop}\right].
    }
\end{equation}
The quantity \(\Delta_\chi\) is a vacuum chiral-channel strength, not an external scattering factor.

\section{The Pure-Glue Topological Stiffness in the Halpern Representation}
\label{sec:pure_glue_halpern}

The second quantity entering the susceptibility formula is \(A_{\rm F}=\chi_{\rm YM}^{\rm F}\). Formally,
\begin{equation}
    A_{\rm F}
    =
    \int d^4x\,
    \langle Q(x)Q(0)\rangle_{{\rm YM},{\rm F}}.
\end{equation}
In the Halpern representation, the CP-odd insertion is represented by the local pseudoscalar invariant \(\chi^a_{\mu\nu}\widetilde\chi^{a\mu\nu}\). Introducing \(\chi^a_{\mu\nu}=\Lambda_{\rm EL}^2y^a_{\mu\nu}\), the local CP-odd charge is
\begin{equation}
    q_\chi(y)=\frac{g^2}{32\pi^2}y^a_{\mu\nu}\widetilde y^{a\mu\nu}.
\end{equation}
For a CP-even measure,
\begin{equation}
    A_{\rm F}=\Lambda_{\rm EL}^4\langle q_\chi^2\rangle_{\rm EL}.
\end{equation}
Decomposing \(y\) into self-dual and anti-self-dual parts gives
\begin{equation}
    y^a_{\mu\nu}\widetilde y^{a\mu\nu}=X_+-X_-,
\end{equation}
and therefore
\begin{equation}
    \boxed{
    A_{\rm F}
    =
    \Lambda_{\rm EL}^4
    \left(\frac{g^2}{32\pi^2}\right)^2
    {\rm Var}_{\rm EL}(X_+-X_-)
    =
    \Lambda_{\rm EL}^4{\cal A}_{\rm F}.
    }
\end{equation}

A regulated local measure may be written schematically as
\begin{equation}
    d\mu_{\rm EL}^{(\xi)}(y)
    =
    \frac{1}{Z_\xi}
    dy\,e^{-y^2/4}
    \left[\det({\cal F}^\dagger{\cal F}+\xi^2{\bf 1})\right]^{-1/4}
    {\cal R}_\xi(y).
\end{equation}
For fixed \(\xi>0\), the variance is finite under mild conditions on \({\cal R}_\xi\). The regulator-independent statement requires the existence of the limit \({\cal A}_{\rm F}=\lim_{\xi\to0^+}{\cal A}_{\rm F}^{(\xi)}\), a nontrivial angular determinant problem in the full Fried formalism.

Combining \(A_{\rm F}=\Lambda_{\rm EL}^4{\cal A}_{\rm F}\) with the chiral condensate gives
\begin{equation}
    \boxed{
    m_a^2f_a^2
    =
    \left[
    \frac{1}{\Lambda_{\rm EL}^4{\cal A}_{\rm F}}
    +
    \frac{4\pi^2}
    {N_cr\Lambda_{\rm EL}^3I(r)}
    \sum_f\frac{1}{m_f}
    \right]^{-1}.
    }
\end{equation}

\section{Relation to the \texorpdfstring{\(U(1)_A\)}{U(1)A} Sector and the \texorpdfstring{\(\eta'\)}{eta prime}}
\label{sec:eta_prime}

The pure-glue stiffness \(A_{\rm F}\) also controls the anomalous singlet pseudoscalar sector. For \(N_f\) approximately degenerate light flavors, write
\begin{equation}
    \phi_f=\frac{\eta_0}{\sqrt{N_f}f_\pi}.
\end{equation}
Then
\begin{equation}
    \sum_f\phi_f=\frac{\sqrt{N_f}\eta_0}{f_\pi}.
\end{equation}
At \(\Theta=0\), the anomalous term gives a singlet mass contribution of order
\begin{equation}
    m_{\eta'}^2
    \sim
    \frac{2N_fA_{\rm F}}{f_\pi^2}
    +
    \text{quark-mass corrections},
\end{equation}
up to normalization conventions. This is the Fried-formalism analogue of the Witten--Veneziano structure~\cite{WittenU1,VenezianoU1}. Thus the same \(A_{\rm F}\) controls the pure Yang--Mills topological susceptibility, the anomalous singlet pseudoscalar mass contribution, and the pure-glue input to the QCD axion potential.

\section{Phenomenological Normalization}
\label{sec:phenomenological_normalization}

In the physical light-quark regime,
\begin{equation}
    m_a^2f_a^2\simeq m_*\Sigma_{\rm F}.
\end{equation}
The standard zero-temperature QCD axion normalization is commonly written as~\cite{GrilliDiCortona,GorghettoVilladoro}
\begin{equation}
    m_a
    \simeq
    5.7~\mu{\rm eV}
    \left(\frac{10^{12}~{\rm GeV}}{f_a}\right),
\end{equation}
corresponding to \(\chi_{\rm top}^{1/4}(0)\simeq75~{\rm MeV}\). Reproducing this value in the light-quark regime requires a condensate scale
\begin{equation}
    \Sigma_{\rm F}^{1/3}\sim270\text{--}285~{\rm MeV},
\end{equation}
which is an ordinary nonperturbative QCD scale. Using
\begin{equation}
    \Sigma_{\rm F}=\frac{N_cr\Lambda_{\rm EL}^3}{4\pi^2}I(r),
\end{equation}
representative values \(r\sim0.4\text{--}0.6\) and \(I(r)=O(1)\) give \(\Lambda_{\rm EL}\sim0.8\text{--}1.0~{\rm GeV}\) and \(M_0\sim0.35\text{--}0.55~{\rm GeV}\).

\section{Several Axion Fields and the Rank-One QCD Contribution}
\label{sec:multi_axion_rank_one}

Let \(a_i\), \(i=1,\ldots,N\), be pseudoscalar fields with decay constants \(f_i\), coupled to QCD through
\begin{equation}
    \Theta_Q=\theta_Q+\sum_i c_i\frac{a_i}{f_i}.
\end{equation}
The Fried-QCD contribution to the potential is
\begin{equation}
    V_Q^{\rm F}=\chi_Q^{\rm F}\left[1-\cos\Theta_Q\right].
\end{equation}
Expanding about the minimum gives
\begin{equation}
    (M_Q^2)_{ij}
    =
    \boxed{
    \chi_Q^{\rm F}\frac{c_ic_j}{f_if_j}.
    }
\end{equation}
This is an outer product and therefore has rank one. Fried-QCD generates one massive QCD axion direction and \(N-1\) QCD-flat directions. Additional massive axion-like particles require additional independent topological sectors.

\section{Finite Temperature and Cosmological Use}
\label{sec:finite_temperature}

The natural finite-temperature extension is
\begin{equation}
    \boxed{
    \chi_{\rm top}^{\rm F}(T)
    =
    \left[
    \frac{1}{A_{\rm F}(T)}
    +
    \sum_f
    \frac{1}{m_f\Sigma_{\rm F}(T)}
    \right]^{-1}.
    }
\end{equation}
Consequently,
\begin{equation}
    m_a^2(T)f_a^2=\chi_{\rm top}^{\rm F}(T).
\end{equation}
The two Fried quantities are now temperature-dependent: \(\Sigma_{\rm F}\to\Sigma_{\rm F}(T)\) and \(A_{\rm F}\to A_{\rm F}(T)\). In the chiral sector,
\begin{equation}
    \Sigma_{\rm F}(T)
    =
    4N_cT
    \sum_n
    \int\frac{d^3p}{(2\pi)^3}
    \frac{M_{\rm F}(\omega_n^2+\mathbf p^2;T)}{\omega_n^2+\mathbf p^2+M_{\rm F}^2(\omega_n^2+\mathbf p^2;T)}.
\end{equation}
In the pure-glue sector,
\begin{equation}
    A_{\rm F}(T)=\Lambda_{\rm EL}^4(T){\cal A}_{\rm F}(T),
    \qquad
    {\cal A}_{\rm F}(T)=\left(\frac{g^2}{32\pi^2}\right)^2{\rm Var}_{{\rm EL},T}(X_+-X_-).
\end{equation}

Coherent axion oscillations begin when
\begin{equation}
    3H(T_{\rm osc})\simeq m_a(T_{\rm osc}).
\end{equation}
During radiation domination, \(H(T)=1.66\sqrt{g_*(T)}T^2/M_{\rm Pl}\). Thus
\begin{equation}
    3\cdot1.66\sqrt{g_*(T_{\rm osc})}\frac{T_{\rm osc}^2}{M_{\rm Pl}}
    =
    \frac{\sqrt{\chi_{\rm top}^{\rm F}(T_{\rm osc})}}{f_a}.
\end{equation}
This shows that the Fried-formalism contribution to axion cosmology is entirely through the finite-temperature susceptibility.

\section{Relation to Standard Treatments of the QCD Axion}
\label{sec:relation_standard}

The standard axion relation \(m_a^2f_a^2=\chi_{\rm top}\) is unchanged. The Fried-specific contribution is the representation
\begin{equation}
    \chi_{\rm top}^{\rm F}
    =
    \left[
    \frac{1}{A_{\rm F}}
    +
    \sum_f
    \frac{1}{m_f\Sigma_{\rm F}}
    \right]^{-1}.
\end{equation}
Thus the nonstandard part is the organization of \(\chi_{\rm top}\) through the chiral Effective Locality quantity \(\Sigma_{\rm F}\) and the pure-glue Halpern quantity \(A_{\rm F}\). This should be read as a bridge between the standard axion relation and the Effective Locality representation of QCD topology.

\section{Required Fried-Formalism Calculations}
\label{sec:required_fried_calculations}

The conditional result becomes a first-principles Fried axion derivation if the following two chains are completed:
\begin{equation}
    \boxed{K_{\rm EL}\longrightarrow G_\chi\longrightarrow \Sigma_{\rm F},}
\end{equation}
and
\begin{equation}
    \boxed{d\mu_{\rm EL}(\chi)\longrightarrow {\rm Var}_{\rm EL}(X_+-X_-)\longrightarrow A_{\rm F}.}
\end{equation}
The first is the scalar/pseudoscalar Effective Locality projection. The second is the CP-odd Halpern angular variance. A successful evaluation of \(A_{\rm F}\) should also reproduce the scale of the anomalous \(U(1)_A\) sector.

\section{Dimensional Analysis and Internal Consistency}
\label{sec:dimensional_consistency}

The dimensions are
\begin{equation}
    [\Sigma_{\rm F}]={\rm mass}^3,
    \qquad
    [A_{\rm F}]={\rm mass}^4,
    \qquad
    [m_f]={\rm mass}.
\end{equation}
Thus \([m_f\Sigma_{\rm F}]={\rm mass}^4\), and the interpolation formula has \([\chi_{\rm top}^{\rm F}]={\rm mass}^4\), as required.

Define
\begin{equation}
    \epsilon_A=\frac{m_*\Sigma_{\rm F}}{A_{\rm F}}.
\end{equation}
Using the Fried expressions,
\begin{equation}
    \epsilon_A
    =
    \frac{N_crI(r)}{4\pi^2{\cal A}_{\rm F}}
    \frac{m_*}{\Lambda_{\rm EL}}.
\end{equation}
For physical light quarks, \(m_*\ll\Lambda_{\rm EL}\), so \(\epsilon_A\) is naturally small unless \({\cal A}_{\rm F}\) is anomalously suppressed. Therefore
\begin{equation}
    \chi_{\rm top}^{\rm F}
    =
    m_*\Sigma_{\rm F}
    \left[1-\epsilon_A+O(\epsilon_A^2)\right].
\end{equation}

\section{Claims and Limitations}
\label{sec:claims_limitations}

The result should be read as a reduction theorem rather than as a completed numerical prediction. The standard axion relation \(m_a^2f_a^2=\chi_{\rm top}\) is unchanged. What is changed is the organization of \(\chi_{\rm top}\): in the Fried framework it is expressed through a chiral Effective Locality quantity \(\Sigma_{\rm F}\) and a pure-glue Halpern quantity \(A_{\rm F}\).

We should not claim that Effective Locality by itself predicts the existence of an axion. The axion field is introduced through the usual Peccei--Quinn mechanism~\cite{PecceiQuinnPRL,PecceiQuinnPRD,WeinbergAxion,WilczekAxion} or an equivalent dynamical \(\Theta\)-angle. We should not claim that Fried-QCD alone produces several massive axion species. The QCD topological density is a single operator, and its contribution to any multi-axion mass matrix is rank one. We should not claim that the present result replaces lattice QCD or chiral perturbation theory. The result is analytic and structural: it identifies how the QCD susceptibility would be represented inside the Fried Effective Locality framework.

The main quantities remaining to be computed from the full measure are \(\Sigma_{\rm F}\) and \(A_{\rm F}\). If these are evaluated directly from the full Fried--Gabellini--Grandou--Tsang--Sheu formalism, the conditional construction developed here would become a genuine Fried-formalism derivation of the QCD axion potential.

\section{Conclusion}
\label{sec:conclusion}

We have examined how the QCD axion potential is represented in the context of Fried's nonperturbative QCD functional formalism. The axion field was introduced in the standard way, through the replacement
\begin{equation}
    \theta_{\rm QCD}\longrightarrow \Theta=\theta_{\rm QCD}+\frac{a}{f_a}.
\end{equation}
The central result is the conditional Fried-formalism representation
\begin{equation}
    \boxed{
    m_a^2f_a^2
    =
    \left[
    \frac{1}{A_{\rm F}}
    +
    \sum_f\frac{1}{m_f\Sigma_{\rm F}}
    \right]^{-1},
    }
\end{equation}
where \(\Sigma_{\rm F}\) and \(A_{\rm F}\) are the chiral and pure-glue Fried-formalism inputs that remain to be evaluated from the full Fried--Gabellini--Grandou--Tsang--Sheu measure.

In the scalar/pseudoscalar projection of the Effective Locality kernel,
\begin{equation}
    \Sigma_{\rm F}=\frac{N_cr\Lambda_{\rm EL}^3}{4\pi^2}I(r),
    \qquad
    1=\alpha_\chi^{\rm F}J(r).
\end{equation}
In the pure-glue Halpern sector,
\begin{equation}
    A_{\rm F}
    =
    \Lambda_{\rm EL}^4
    \left(\frac{g^2}{32\pi^2}\right)^2
    {\rm Var}_{\rm EL}(X_+-X_-).
\end{equation}
Thus Fried's nonperturbative QCD formalism, through the property of Effective Locality, gives a QCD-first route to the axion potential by reducing it to \(\Sigma_{\rm F}\) and \(A_{\rm F}\).

If these two quantities can be evaluated directly from the full Fried--Gabellini--Grandou--Tsang--Sheu functional measure, then the QCD axion mass, the pure Yang--Mills topological susceptibility, and the anomalous \(U(1)_A\) singlet sector would be controlled by the same nonperturbative structures underlying Fried's formalism.

\appendix

\section{Fierz Projection of the Color Current}
\label{app:fierz_projection}

We begin from
\begin{equation}
    {\cal L}_{\rm EL}^{(4q)}
    =
    -G_{\rm EL}
    \left(\bar q\gamma_\mu T^a q\right)
    \left(\bar q\gamma^\mu T^a q\right),
\end{equation}
with \({\rm tr}(T^aT^b)=\delta^{ab}/2\). The color identity
\begin{equation}
    T^a_{ij}T^a_{kl}
    =
    \frac12
    \left(\delta_{il}\delta_{jk}-\frac1{N_c}\delta_{ij}\delta_{kl}\right)
\end{equation}
gives the color-singlet coefficient
\begin{equation}
    C_\chi=\frac{N_c^2-1}{2N_c^2}.
\end{equation}
For QCD, \(C_\chi=4/9\). Different normalizations redistribute this factor between \(C_\chi\) and \(G_{\rm EL}\); the physical object is \(G_\chi=C_\chi G_{\rm EL}\).

\section{Gap-Equation Integrals}
\label{app:gap_integrals}

The scalar/pseudoscalar approximation uses
\begin{equation}
    J(r)=\int_0^\infty dx\,\frac{x e^{-2x}}{x+r^2e^{-2x}},
    \qquad
    I(r)=\int_0^\infty dx\,\frac{x e^{-x}}{x+r^2e^{-2x}}.
\end{equation}
Since \(J(0)=1/2\), the critical value is \(\alpha_{\chi,c}^{\rm F}=2\). Representative values are
\begin{equation}
\begin{array}{c|c|c|c}
r & I(r) & J(r) & \alpha_\chi^{\rm F}=1/J(r)\\
\hline
0.3 & 0.88 & 0.39 & 2.6\\
0.4 & 0.83 & 0.36 & 2.8\\
0.5 & 0.79 & 0.33 & 3.0\\
0.6 & 0.75 & 0.30 & 3.3\\
0.8 & 0.69 & 0.26 & 3.9\\
1.0 & 0.64 & 0.23 & 4.4
\end{array}
\end{equation}

\section{Self-Dual Halpern Variables and the CP-Odd Variance}
\label{app:self_dual_halpern}

In Euclidean signature define
\begin{equation}
    y_{\mu\nu}^{a,\pm}=\frac12\left(y^a_{\mu\nu}\pm\widetilde y^a_{\mu\nu}\right).
\end{equation}
Then
\begin{equation}
    y^a_{\mu\nu}\widetilde y^{a\mu\nu}=X_+-X_-.
\end{equation}
For \(SU(3)\), each of the self-dual and anti-self-dual sectors has \(3(N_c^2-1)=24\) components. A Gaussian reference measure gives \({\rm Var}_0(X_+-X_-)=384\), but the full Fried value is modified by determinant and angular factors.

\section{The \texorpdfstring{\(\Theta\)}{Theta}-Deformed Halpern Identity}
\label{app:theta_halpern_identity}

The Euclidean pure-glue action with source is
\begin{equation}
    S_{\rm YM,\Theta}
    =
    \frac14\int G^2
    -
    i\Theta\frac{g^2}{32\pi^2}\int G\widetilde G.
\end{equation}
Formally, the Halpern identity generalizes to
\begin{equation}
    \exp\left[-\frac14\int G{\cal M}_\Theta G\right]
    =
    {\cal N}_\Theta
    \int{\cal D}\chi\,
    \exp\left[-\frac14\int\chi{\cal M}_\Theta^{-1}\chi+\frac{i}{2}\int\chi G\right],
\end{equation}
where \({\cal M}_\Theta=1-i\Theta c_\theta\star\). Expanding in \(\Theta\) shows that the first \(\Theta\)-derivative inserts \(\chi\widetilde\chi\). Thus \(G\widetilde G\leadsto\chi\widetilde\chi\) is not a classical pointwise equality; it is the representation of the CP-odd source insertion after the Halpern transformation.

\section{Euclidean Conventions and Normalization of the Topological Term}
\label{app:euclidean_conventions}

In Euclidean signature,
\begin{equation}
    Z_E(\theta)=\int{\cal D}A\,{\cal D}q\,{\cal D}\bar q\,
    \exp\left[-S_E+i\theta\int d^4x\,Q_E(x)\right].
\end{equation}
The Euclidean vacuum energy is \(E_E(\theta)=-V_4^{-1}\ln Z_E(\theta)\), and
\begin{equation}
    \chi_{\rm top}=\left.\frac{\partial^2E_E}{\partial\theta^2}\right|_{\theta=0}.
\end{equation}
At \(\theta=0\), this is the variance of the integrated topological charge divided by \(V_4\), and is positive for a positive Euclidean measure. This convention gives \(m_a^2f_a^2=\chi_{\rm top}\).

\bibliographystyle{apsrev4-2}
\bibliography{fried_axion_fried_formalism_names_refs}

@article{FriedReview,
  author        = {Fried, H. M. and Gabellini, Y. and Grandou, T. and Tsang, P. H.},
  title         = {{QCD Effective Locality: A Theoretical and Phenomenological Review}},
  journal       = {Universe},
  volume        = {7},
  pages         = {481},
  year          = {2021},
  eprint        = {2111.09221},
  archivePrefix = {arXiv},
  primaryClass  = {hep-ph}
}

@article{FriedPureGlue,
  author        = {Fried, H. M. and Gabellini, Y. and Grandou, T.},
  title         = {{Effective Locality in the Pure Gluon Sector}},
  journal       = {arXiv e-prints},
  pages         = {arXiv:2203.16456},
  year          = {2022},
  eprint        = {2203.16456},
  archivePrefix = {arXiv},
  primaryClass  = {hep-th}
}

@article{PecceiQuinnPRL,
  author  = {Peccei, R. D. and Quinn, H. R.},
  title   = {{CP Conservation in the Presence of Pseudoparticles}},
  journal = {Phys. Rev. Lett.},
  volume  = {38},
  pages   = {1440--1443},
  year    = {1977}
}

@article{PecceiQuinnPRD,
  author  = {Peccei, R. D. and Quinn, H. R.},
  title   = {{Constraints Imposed by CP Conservation in the Presence of Pseudoparticles}},
  journal = {Phys. Rev. D},
  volume  = {16},
  pages   = {1791--1797},
  year    = {1977}
}

@article{WeinbergAxion,
  author  = {Weinberg, Steven},
  title   = {{A New Light Boson?}},
  journal = {Phys. Rev. Lett.},
  volume  = {40},
  pages   = {223--226},
  year    = {1978}
}

@article{WilczekAxion,
  author  = {Wilczek, Frank},
  title   = {{Problem of Strong P and T Invariance in the Presence of Instantons}},
  journal = {Phys. Rev. Lett.},
  volume  = {40},
  pages   = {279--282},
  year    = {1978}
}

@article{WittenU1,
  author  = {Witten, Edward},
  title   = {{Current Algebra Theorems for the U(1) Goldstone Boson}},
  journal = {Nucl. Phys. B},
  volume  = {156},
  pages   = {269--283},
  year    = {1979}
}

@article{VenezianoU1,
  author  = {Veneziano, G.},
  title   = {{U(1) Without Instantons}},
  journal = {Nucl. Phys. B},
  volume  = {159},
  pages   = {213--224},
  year    = {1979}
}

@article{GrilliDiCortona,
  author        = {di Cortona, Giovanni Grilli and Hardy, Edward and Pardo Vega, Javier and Villadoro, Giovanni},
  title         = {{The QCD Axion, Precisely}},
  journal       = {JHEP},
  volume        = {2016},
  number        = {01},
  pages         = {034},
  year          = {2016},
  eprint        = {1511.02867},
  archivePrefix = {arXiv},
  primaryClass  = {hep-ph}
}

@article{GorghettoVilladoro,
  author        = {Gorghetto, Marco and Villadoro, Giovanni},
  title         = {{Topological Susceptibility and QCD Axion Mass: QED and NNLO Corrections}},
  journal       = {JHEP},
  volume        = {2019},
  number        = {03},
  pages         = {033},
  year          = {2019},
  eprint        = {1812.01008},
  archivePrefix = {arXiv},
  primaryClass  = {hep-ph}
}

@article{BorsanyiAxion,
  author        = {Borsanyi, S. and others},
  title         = {{Calculation of the Axion Mass Based on High-Temperature Lattice Quantum Chromodynamics}},
  journal       = {Nature},
  volume        = {539},
  pages         = {69--71},
  year          = {2016},
  eprint        = {1606.07494},
  archivePrefix = {arXiv},
  primaryClass  = {hep-lat}
}

\end{document}